\documentclass[a4paper,11pt]{article}
\usepackage{jheppub}
\usepackage{amsmath}
\usepackage{amsfonts}
\usepackage{amssymb}
\usepackage{textcomp}
\usepackage{graphicx}
\usepackage{bm}
\usepackage{color}
\usepackage{verbatim}
\usepackage{subfig}
\usepackage{graphicx,color}
\usepackage{epsfig}

\usepackage{amsfonts,amsmath,amssymb}

\renewcommand{\Im}{\operatorname{Im}}
\renewcommand{\Re}{\operatorname{Re}}

\begin{document}

 \title{Drude in D major}
 
 \author[a]{Tom\'as Andrade,}
 \affiliation[a]{Rudolf Peierls Centre for Theoretical Physics, University of Oxford, 1 Keble Road, Oxford OX1 3NP, UK} 
  
\author[b]{Simon A. Gentle}
\affiliation[b]{Department of Physics and Astronomy, University of California, Los Angeles, CA 90095, USA}

\author[c,d]{and\ Benjamin Withers}
\affiliation[c]{School of Mathematical Sciences, Queen Mary University of London, Mile End Road, London E1 4NS, UK}
\affiliation[d]{DAMTP, Centre for Mathematical Sciences, Wilberforce Road, Cambridge CB3 0WA, UK} 

\emailAdd{tomas.andrade@physics.ox.ac.uk}
\emailAdd{sgentle@physics.ucla.edu}
\emailAdd{b.withers@qmul.ac.uk}

\abstract{We study holographic momentum relaxation in the limit of a large number of spacetime dimensions $D$. 
For an axion model we find that momentum conservation is restored as $D$ becomes large.
To compensate we scale the strength of the sources with $D$ so that momentum is relaxed even at infinite $D$. 
We analytically obtain the quasi-normal modes which control electric and heat transport, and give their frequencies in a $1/D$ expansion. 
We also obtain the AC thermal conductivity as an expansion in $1/D$, which at leading order takes Drude form. 
To order $1/D$ our analytical result provides a reasonable approximation to the AC conductivity even at $D=4$,
establishing large $D$ as a practical method in this context.
As a further application, we discuss the signature of the transition from coherent to incoherent behaviour known to exist in the system for finite $D$.
}

\maketitle
 
\section{Introduction}
\label{sec:intro}
\setcounter{page}{2}

An interesting application of gauge/gravity duality to condensed matter physics arises in the study of 
momentum relaxation. This is so mainly because the resulting zero frequency conductivities are finite, 
allowing us to study transport in a more realistic way.
To this end, one must find gravitational solutions which break translational 
invariance along the boundary directions due to the presence of one or more spatially dependent sources. 
Generically this involves numerically solving the Einstein equations which give an elliptic PDE problem in this context.
Some studies focus on configurations which describe a lattice in the dual field theory, for instance when the chemical potential is a periodic function 
of a spatial direction on the boundary.
Gravitational solutions of this kind have been successfully constructed 
\cite{Horowitz:2012ky, Horowitz:2012gs, Horowitz:2013jaa, Donos:2014yya, Rangamani:2015hka}
and they indeed reproduce the expected
low frequency dynamics, i.e.\ the zero frequency delta functions in the conductivities are resolved into finite width
Drude peaks.   

A considerable technical simplification arises if a certain global symmetry is present, say, in the matter sector 
of the bulk theory. Then, translational invariance can be broken along that symmetry direction while 
preserving homogeneity of the geometry, which in turn 
implies that the construction of such solutions only requires solving ODEs. Examples of this kind include
\cite{Donos:2012js, Donos:2013eha}, which have been shown to yield finite DC conductivities as well as to possess a rich 
structure which displays transitions between different metallic and 
insulating regimes.\footnote{A simplification of this type arises in holography with massive gravity in the bulk \cite{Vegh:2013sk}.}

We can  simplify the problem even further by arranging the bulk matter content in such a way that the 
black branes of interest are not only homogeneous but also isotropic. This fact was exploited in \cite{Andrade:2013gsa} 
(see also \cite{Taylor:2014tka}), which considered a particular configuration of a set of massless scalars, termed linear axions since they 
are shift-symmetric, that allows for an analytical black brane 
solution at non-zero chemical potential in an arbitrary number of dimensions. As expected, the DC 
electric and thermal conductivities are finite \cite{Andrade:2013gsa, Gouteraux:2014hca,Davison:2014lua,Donos:2014cya}, and they can be evaluated 
analytically. However, the conductivities at non-zero frequency need to be computed numerically since this 
involves solving coupled fluctuation equations around the background solution. Interestingly, it has been noted that 
deviations from Drude physics are present when the strength of momentum relaxation is large 
\cite{Kim:2014bza, Davison:2014lua, Davison:2015bea}.

Recently, it has been noted that the equations of General Relativity simplify considerably in the limit in 
which the number of spacetime dimensions, $D$, is taken to be large, which provides an efficient tool to 
approximate finite $D$ results, as a perturbative calculation in $1/D$  \cite{Emparan:2013moa}. 
The main ingredient of the construction is the fact that when the number of spacetime dimensions is large, 
the gravitational potential becomes very steep near the horizon which yields to a natural separation of the 
dynamics that localise near the horizon and the ones that probe regions far away from it. 
In particular, this implies that the quasi-normal mode (QNM) spectrum splits into near-horizon {\it decoupled} modes,
and {\it coupled} modes that are delocalised \cite{Emparan:2014cia, Emparan:2014aba, Emparan:2015rva}. 
As explained in these references, while the coupled modes are quite generic, i.e.\ 
shared by many black holes, the decoupled modes are sensitive to the particularities of different 
solutions. 

In this paper we initiate the study of holographic inhomogeneities and the resulting momentum relaxation using large $D$ techniques.
Our first goal is to study the decoupled quasi-normal modes which control the characteristic decay rate of the  
electric and thermal conductivities in the linear axion model at non-zero chemical potential. 
We find that momentum conservation is restored unless we scale the strength of the axions with $D$, i.e.\ the decoupled QNM frequency vanishes to leading order. 
Therefore we scale the axion strength appropriately and obtain a QNM capturing the momentum decay rate at leading order in large $D$. We calculate corrections to this in $1/D$.
We see that in certain regimes the QNM frequencies are well-described using the large $D$ approximation. 
Second, we will compute the AC thermal conductivity for zero chemical potential, which at leading order is exactly of Drude form. These results are consistent with the corresponding QNM calculation. To our knowledge, this is the first 
analytical realisation of Drude behaviour outside of the hydrodynamic regime in the context of holography.
With these results at hand, we will comment on the signature of the transition from coherent to incoherent regimes, 
i.e.\ the breakdown of Drude physics, in the large $D$ approximation.   

This paper is organised as follows. In section \ref{sec:model} we review the axion model, its transport properties and the appropriate master fields for the conductivity calculation. In section \ref{sec:QNM} we compute the large $D$ decoupled QNMs which control momentum relaxation, giving analytical expressions for their frequencies. In section \ref{sec:conductivity} we compute the AC thermal conductivity as an expansion in $1/D$. We conclude in section \ref{sec:conclusions}.

\section{Momentum relaxation in arbitrary D}
\label{sec:model}

In this section we review the holographic model of momentum relaxation proposed in~\cite{Andrade:2013gsa}. 
We will discuss the main properties of the background solution and describe the computation of the two-point 
functions using a gauge invariant master field formalism.

\subsection{Linear axion background}
\label{sec:axion}

The holographic model of momentum relaxation in $D = n+3$ bulk dimensions which we consider throughout this 
paper is given by the action \cite{Andrade:2013gsa}
\begin{equation}\label{S0}
	S_0 = \int d^{n+3} x \sqrt{- g} \left( R + (n+1)(n+2) \ell^{-2} - \frac{1}{4} F^2 - \frac{1}{2} \sum_{I=1}^{n+1} (\partial \psi_I)^2  \right)
\end{equation}
\noindent where $F = dA$ is the field strength of a $U(1)$ gauge field, $\psi_I$ are $(n+1)$ massless scalar fields 
and $\ell$ is the AdS radius which we set to one henceforth. 

This model admits the following analytical black brane solution\footnote{This solution was previously derived in \cite{Bardoux:2012aw} 
in a different context.}
\begin{equation}\label{NL ansatz}
	ds^2 = - f(r) dt^2 + \frac{dr^2}{f(r)} + r^2 \delta_{a b} dx^a dx^b, \qquad A = A_t(r) dt, \qquad \psi_I = \delta_{I a} x^a
\end{equation}
\noindent where $a$ labels the $(n+1)$ boundary spatial directions $x^a$ and
\begin{align}
	f(r) &= r^2 - \frac{\alpha^2}{2 n} - \frac{m_0}{r^n} + \frac{n \mu^2}{2(n+1)} \frac{r_0^{2n}}{r^{2n}}\label{axion soln} \\
	A_t(r) &= \mu \left(1  - \frac{r_0^n}{r^n} \right)\label{axion gauge field}
\end{align}
Here $r_0$ is the horizon of the brane, $\mu$ is the chemical potential in the dual theory and $m_0$ is related to the total energy
of the solution. 
The Hawking temperature is given by 
\begin{equation}
	T = \frac{f'(r_0)}{4 \pi} = \frac{1}{4 \pi} \left( (n+2) r_0 - \frac{\alpha^2}{2 r_0} - \frac{n^2 \mu^2}{2 (n+1) r_0} \right).
\end{equation}
Note that, despite the fact that the geometry is isotropic and homogeneous, the solution manifestly breaks translational
invariance due to the explicit dependence of $\psi_I$ on $x^a$. This feature is reflected in the its thermoelectric DC conductivities, 
with the $\delta$-function at zero frequency present in the Reissner-Nordstr\"om solution removed due to the breaking of translational invariance. 

\subsection{Transport}
\label{sec:Trans}
Conductivities can be computed in terms of two-point functions, which are given in AdS/CFT by studying linear fluctuations
around the black holes under consideration.  Here we are interested in the electric and thermal conductivities at zero spatial 
momentum, which can be obtained in terms of the retarded two-point functions
\begin{equation}\label{2pt fns}
	G_{JJ} (\omega) = \langle J^1 J^1 \rangle (\omega), \quad G_{QJ} (\omega) = \langle Q^1J^1 \rangle (\omega) \quad G_{QQ} (\omega) = \langle Q^1Q^1 \rangle (\omega)
\end{equation}
\noindent where $Q^i = T^{ti} - \mu J^i$ and $T^{ij}$ and $J^i$ are the the stress tensor and $U(1)$ current of the field theory, respectively. 
Here we have chosen to compute the conductivities along the axis $x^1$. Because the black holes of interest are isotropic, this does not 
result in loss of generality. 
We can then express the electric conductivity $\sigma(\omega)$, the thermo-electric conductivity $\beta(\omega)$ and the thermal conductivity $\kappa(\omega)$ 
in terms of the two-point functions \eqref{2pt fns} by means of the Kubo formulae:
\begin{align}
\nonumber
	\sigma(\omega) &= \frac{i}{\omega} (G_{JJ} (\omega) - G_{JJ} (0))  , \\
\nonumber
	\beta(\omega)   &= \frac{i}{\omega T} (G_{QJ} (\omega) - G_{QJ} (0))  , \\ 
\label{conductivityDefs}
	\kappa(\omega) &= \frac{i}{\omega T} ( G_{QQ} (\omega) - G_{QQ} (0) )  
\end{align}

Analytical traction may be gained in the DC limit, where these conductivities can be computed. 
As shown in \cite{Andrade:2013gsa} the DC electrical conductivity is given by
\begin{equation}
	\sigma(0) = r_0^{n-1} \left(1 + n^2\frac{\mu^2}{\alpha^2} \right)
\end{equation}
whilst the thermal and thermo-electric conductivities for general $n$ are given in \cite{Donos:2014cya}
\begin{equation}
	\kappa(0) =  r_0^{n+1}\frac{(4\pi)^2 T}{\alpha^2}, \qquad \beta(0) =  r_0^{n}\frac{4\pi \mu}{\alpha^2}. \label{kappaDC} 
\end{equation}

Separately, the conductivities may be approximated analytically for small $\alpha$ by the Drude formula. 
For instance, at $n=1$ the thermal conductivity is given by 
\begin{equation}\label{drude}
	\kappa(\omega) = \frac{\kappa(0)}{1 - i \omega \tau}, \qquad \omega \ll T
\end{equation}
where $\tau$ is the characteristic time of momentum relaxation, set by $\alpha$. Since there is only 
one characteristic time scale, we say that transport is {\it coherent} in this regime.\footnote{See \cite{Hartnoll:2014lpa} for a discussion on this terminology.}
Increasing $\alpha$, the deviations from \eqref{drude} become large, driving the system into an {\it incoherent} 
phase. This transition was first observed in this holographic system by a numerical analysis in $n=1$ \cite{Kim:2014bza}, 
and later on also noticed in the presence of a charged scalar condensate in \cite{Andrade:2014xca}. A closely related coherent/incoherent
transition has been reported for the thermal conductivity at zero chemical potential for $n=1$ in \cite{Davison:2014lua}, which 
focussed on an explanation in terms of QNM: the system behaves coherently when there is an isolated, long-lived, 
purely dissipative excitation in the spectrum. Moreover, this analysis was extended in perturbation theory to include 
chemical potential \cite{Davison:2015bea}, with qualitatively similar results.

\subsection{Master fields}
\label{sec:MF}
A general approach to computing the conductivities in the background \eqref{NL ansatz}-\eqref{axion gauge field} utilises a minimal, consistent set of perturbations,
\begin{equation}\label{linear perts}
  \delta A = e^{- i \omega t} a(r) dx^1, \qquad 	\delta (ds^2) = 2 e^{- i \omega t} r^2 h(r) dt dx^1 , 
  \qquad \delta \psi_1 = e^{- i \omega t} \alpha^{-1} \chi(r).
\end{equation}
The linearised equations of motion which govern the perturbations \eqref{linear perts} can be written as
\begin{align}
	a'' + \left[ \frac{f'}{f} + \frac{(n-1)}{r} \right] a' + \frac{\omega^2}{f^2} a + \frac{\mu n}{f} \frac{r_0^n}{r^{n-1}} h' &= 0 \\
	\chi'' + \left[ \frac{f'}{f} + \frac{(n+1)}{r} \right] \chi' + \frac{\omega^2}{f^2} \chi - \frac{i \omega \alpha^2}{f^2} h &=0 \\
	\frac{i \omega r^2}{f} h' + \frac{i \omega n \mu}{f} \frac{r_0^n}{r^{n+1}} a - \chi' &=0 
\end{align}
where primes denote derivatives with respect to $r$. For odd $n$, the near boundary expansions for the physical fields are given by 
\begin{align}
\label{UV phys1}
	h &= h^{(0)} + \ldots + \frac{h^{(n+2)}}{r^{n+2}}  	  + \ldots\\
	a &= a^{(0)} + \ldots + \frac{a^{(n)}}{r^n}   + \ldots \\
\label{UV phys3}
	\chi &= \chi^{(0)} + \frac{\chi^{(1)}}{r} + \frac{\chi^{(2)}}{r^2}  + \ldots
\end{align}
For even $n$, the expansions \eqref{UV phys1}-\eqref{UV phys3} contain logarithms, as a result of the Weyl anomaly 
present in even boundary dimensions \cite{Henningson:1998gx}. These terms will play no role in the following, so we 
shall omit them. 
The terms $\chi^{(1)} $, $\chi^{(2)} $ are fixed by the equations of motion as
\begin{equation}\label{chi 1 2}
	\chi^{(1)} = 0 , \qquad \chi^{(2)}  = \frac{\omega( \omega \chi^{(0)} - i \alpha^2 h^{(0)})}{2 n}
\end{equation}
The gauge invariant sources for the electric and thermal conductivity are $a^{(0)}$ and 
$ s^{(0)} =\omega \chi^{(0)} - i \alpha^2 h^{(0)} $, respectively (see e.g.\ \cite{Donos:2013eha}).

As shown in \cite{Andrade:2013gsa}, the perturbation equations can be decoupled in terms of two gauge invariant master 
fields $\Phi_\pm$, given by 
\begin{equation}\label{MF def}
   f  r \chi'  = \frac{\omega}{\mu} ( \tilde c_+ \Phi_+ + \tilde c_- \Phi_-  ),  \qquad
	a = - i (\Phi_+ + \Phi_-)
\end{equation}
\noindent where
\begin{equation}\label{cpm}
	\tilde c_\pm = \frac{1}{2 r_0^n} \left\{ (n+2) m_0 \pm [ (n+2)^2 m_0^2 + 4 r_0^{2n} \mu^2 \alpha^2  ]^{1/2} \right \}.
\end{equation}

\noindent The master fields are governed by the equations  
\begin{equation}\label{MF eqs pm}
	r^{3-n} ( f r^{n-1} \Phi_\pm' )' + \left(  \frac{r^2 \omega^2}{f} - \frac{n^2 \mu^2 r_0^{2n}}{r^{2n}} + 
	n \tilde c_\pm \frac{r_0^n}{r^n} \right) \Phi_\pm  = 0.
\end{equation}
\noindent As shown in \cite{Son:2002sd}, in order to obtain the retarded correlators the fluctuations must satisfy ingoing boundary conditions 
at the black hole horizon. These can be implemented by simply imposing the ingoing condition on the master field 
\cite{Berti:2009kk}, which amounts to
\begin{equation}\label{ingoing bc}
	\Phi_\pm(r) = (r - r_0)^{- i \omega/(4 \pi T)} ( \Phi_\pm^H + \ldots  ), \qquad {\rm near } \; r = r_0
\end{equation}
\noindent where $\Phi_\pm^H$ are arbitrary constants and the ellipsis denotes regular subleading terms. 

The near boundary asymptotics of the master fields are given by 
\begin{equation}\label{UV MF}
	\Phi_\pm = \Phi^{(0)}_\pm + \ldots + \frac{1}{r^{n}} \Phi^{(n)}_\pm + \ldots
\end{equation}
From \eqref{UV phys1}-\eqref{UV phys3} and \eqref{MF def}, we learn that the asymptotic data in \eqref{UV MF} 
is related to the physical asymptotic data as
\begin{align}
	\Phi^{(0)}_\pm &= \pm \frac{1}{\omega(\tilde c_- - \tilde  c_+)} ( 2 \mu \chi^{(2)} + i \omega \tilde c_- a^{(0)}  ) \\
	\Phi^{(n)}_\pm &= \pm \frac{i}{\omega^2(\tilde c_- - \tilde  c_+)} ( - \alpha^2 (n+2) \mu h^{(n+2)} + \omega^2 \tilde c_- a^{(n)}  ) 
\end{align}
\noindent where $\chi^{(2)}$ is related to the gauge invariant source for the stress tensor by \eqref{chi 1 2}.
In order to compute the two-point functions at $\mu\neq 0$, a detailed computation of the on-shell action is needed due to the 
non-trivial interplay between the physical sources and vevs in $\Phi_\pm$\footnote{This computation was carried out
for $n=1$ in \cite{Davison:2015bea}.}. However,  it is easy to see that in order to obtain the poles in such correlators 
it suffices to solve for the spectra of $\Phi_\pm$ with Dirichlet boundary conditions $\Phi^{(0)}_\pm =0$.

\subsubsection{The neutral case}
\label{neutral master}

For $\mu = 0$, all the gauge invariant information is contained in the thermal conductivity. To compute it, the 
relevant fluctuations are \eqref{linear perts} with $a(r) = 0$. The physical boundary data satisfies \eqref{chi 1 2} 
and the gauge invariant source for the stress tensor is again $ s^{(0)}$. 
Via simple manipulations of the equations of motion, we can derive the master field equation
\begin{equation}\label{MF eq neutral}
	r^{3-n} ( f r^{n-1} \Phi' )' + \left(  \frac{r^2 \omega^2}{f}  + n (n+2) \frac{m_0}{r^n} \right) \Phi = 0 
\end{equation}
\noindent where the master field $\Phi$ is given by
\begin{equation}\label{Phi def}
	\Phi =   \frac{f r \chi'}{i\omega}
\end{equation}
Note that \eqref{MF eq neutral} is the $\mu \to 0$ limit of the equation for $\Phi_+$ \eqref{MF eqs pm} with $m_0 \geq 0$.
Equation \eqref{MF eq neutral} has been previously derived
for $n=1$ in \cite{Davison:2014lua}.
As in the $\mu \neq 0$ case, the UV asymptotics for $\Phi$ can be written as
\begin{equation}\label{UV Psi}
	\Phi = \Phi^{(0)} + \ldots + \frac{\Phi^{(n)}}{r^{n}}  + \ldots
\end{equation}
\noindent where once again we are not writing down the terms involving $ \log r $ which are present for 
even $n$. The independent coefficients in \eqref{UV Psi} are related to the boundary data \eqref{UV phys1} and \eqref{UV phys3}
by
\begin{equation}\label{Phi UV data}
	\Phi^{(0)} = \frac{ i\omega \chi^{(0)} + \alpha^2 h^{(0)}}{n}, \qquad 
	\Phi^{(n)} = \frac{ (n+2) \alpha^2}{\omega^2}   h^{(n+2)}
\end{equation}

Up to an overall $\omega$-independent factor, $\xi$, which we will fix later using the DC results, the two-point function $G_{QQ}$ can be written as 
\begin{equation}\label{G2 neutral}
	G_{QQ} = \xi\frac{\Phi^{(n)}}{\Phi^{(0)}} 
\end{equation}
Here we have chosen a renormalization scheme in which all local contributions to \eqref{G2 neutral} are removed by 
counterterms \cite{deHaro:2000vlm}.

\section{QNM frequencies}
\label{sec:QNM}

Finding analytical solutions to the master field equations \eqref{MF eqs pm} and \eqref{MF eq neutral} for general 
values of the parameters seems out of reach. Closely following \cite{Emparan:2013moa, Emparan:2014cia, Emparan:2014aba, 
Emparan:2015rva},  we obtain perturbative 
solutions using $1/n$ as the expansion parameter. 
In this section we will find expressions for the decoupled QNM for $\mu \neq 0$ 
to order $n^{-1}$ and for $\mu = 0 $ to order $n^{-3}$, finding good agreement with numerical calculations at finite $n$
in a certain region of parameter space.  
In  section~\ref{sec:conductivity}  we will carry out the computation of the AC thermal conductivity to  order $n^{-2}$, obtaining a result consistent with our 
QNM calculation.

As explained in \cite{Emparan:2014cia, Emparan:2014aba, Emparan:2015rva}, the spectrum of QNM in the large $n$ limit splits into decoupled 
modes, which are normalisable in the near horizon geometry, and non-decoupled modes,  which are not. The latter 
are shared by many black holes so we do not expect to obtain information about the conductivities in this set of 
modes, since, in particular, they are part of the spectra of black holes which are translationally invariant along 
the boundary directions.
We focus on the decoupled modes and find that they indeed correspond to `Drude poles', i.e.\ they are the purely 
imaginary modes which control the relaxation time of the system. As stated in \cite{Emparan:2014cia, Emparan:2014aba, Emparan:2015rva}, 
a necessary condition for the existence of decoupled QNMs is the presence of negative minima in the effective potential $V_{\pm}$ defined by recasting the master field 
equation as
\begin{equation}
	\left( \frac{d^2}{d r_*^2} + \omega^2 - V_{\pm} \right) \Psi_\pm = 0
\end{equation}
\noindent where $d r_* = dr /f(r)$ is the tortoise coordinate. This form can be achieved by letting $ \Phi_\pm(r) = r^{(1-n)/2} \Psi_\pm(r)$ 
in the master field equations \eqref{MF eqs pm} and \eqref{MF eq neutral}. By examining $V_{-}$, we conclude that there are 
no decoupled QNMs for $\Phi_-$.  

When taking the $n \to \infty$ limit, it is important to assign the scaling with $n$ of different physical 
quantities. Our goal is to capture the effects of momentum relaxation, so we will rescale quantities as appropriate so that $\alpha$ appears at infinite $n$. More concretely, we will 
take the $n \to \infty$ limit holding $r_0$, $\mu$ and $\hat \alpha$ fixed, where
\begin{equation}
  	\hat \alpha = \frac{\alpha}{\sqrt{n}}.
\end{equation}  
This scaling mirrors the scaling of momenta required in \cite{Emparan:2015rva}.
It is convenient to define the radial variable $\rho$ by
\begin{equation}
 	\rho = \left( \frac{r}{r_0} \right)^n.
\end{equation} 
Here we work with $r_0=1$. We will keep $\rho$ finite as $n \to \infty$, performing expansions of
\begin{equation}
r = \rho^{1/n} = 1 + \frac{1}{n}\log{\rho} + \ldots
\end{equation} 
which takes us into the horizon region.
In order to obtain the perturbative solution we are after, we postulate the following expansions for the fields and the 
QNM frequency $\omega = \omega_\pm$,
\begin{equation}\label{n expansions}
	\Phi_\pm(\rho) = \sum_{i = 0} \frac{\Phi_{\pm,i}(\rho)}{n^i}, \qquad \omega_\pm = \sum_{i = 0} \frac{\omega_{\pm,i}}{n^i}
\end{equation}
Our boundary conditions are normalisability in the near horizon, i.e.\ 
\begin{equation}
	\Phi_{\pm,i}(\rho) \to 0 , \qquad {\rm at} \, \, \rho \to \infty 
\end{equation}
\noindent and ingoing boundary conditions at the horizon. These can be written as boundary conditions for the $\Phi_{\pm,i}$
in \eqref{n expansions} by expanding \eqref{ingoing bc} in powers of $1/n$. 
The remainder of the computation of the decoupled QNM proceeds in close parallel to the one described in 
\cite{Emparan:2015rva}, and we shall simply quote our results.

For $\Phi_+$, we find decoupled QNMs with frequencies
\begin{align}
\omega_{+} &= -i \hat\alpha ^2\left\{\frac{ \left(2-\hat\alpha ^2\right)}{2-\hat\alpha ^2+\mu ^2}  - \frac{1}{n}\left[\frac{2 \hat\alpha^2}{\left(2-\hat\alpha ^2+\mu ^2\right) } \,  \log \left(\frac{2-\hat\alpha ^2}{2-\hat\alpha ^2-\mu ^2}\right) \right.\right. \nonumber\\
 &\phantom{=\ }\left.\left. +\frac{2 \left(2 - \hat\alpha^2\right)^3 + \left(12 - 
    8 \hat\alpha^2 + \hat\alpha^4\right) \mu^2 + \left(2 - 3 \hat\alpha^2\right) \mu^4 }{\left(2-\hat\alpha ^2+\mu ^2\right)^3 }  \right]+O\left(n^{-2}\right)\right\}. \label{eq:QNMcharged}
\end{align}
We find that it is impossible to satisfy the  boundary conditions for $\Phi_-$, so we conclude that there are no decoupled QNMs for this field, as argued above. 
In the $\mu=0$ case we are able to obtain two higher orders in the expansion for the $\Phi$ frequency:\footnote{Interestingly, these QNM frequencies can be obtained directly from the QNMs of black branes in AdS without momentum relaxation \cite{Emparan:2015rva} by mapping the momenta $\hat{q}^2 = \hat\alpha^2/2$ and the spatial metric curvature parameter $K=-\hat\alpha^2/2$.}
\begin{align}
\omega &= -i \hat{\alpha}^2  \left\{1 - \frac{2}{n} - \frac{2\left(12+(\pi^2-6)\hat{\alpha}^2\right)}{3\left(\hat{\alpha}^2-2\right)n^2} \right. \nonumber\\
 &\phantom{=\ }\left.+\frac{8\left[-12+\hat{\alpha}^2\left((\hat{\alpha}^2 -4)(\pi^2-3)-3(\hat{\alpha}^2 +2)\zeta(3)\right)\right]}{3\left(\hat{\alpha}^2-2\right)^2 n^3}+O\left(n^{-4}\right)\right\} \label{eq:QNMneutral}.
\end{align} 

A notable feature of the frequencies \eqref{eq:QNMcharged} and \eqref{eq:QNMneutral} is a breakdown of the expansion when $\hat\alpha^2+\mu^2 =2$. In fact this behaviour could have been predicted by examining a large~$n$ expansion of the DC thermal conductivity, \eqref{kappaDC},
\begin{equation}
\kappa(0) = \kappa(0)|_{n\to\infty} \left(1 + \frac{4 + \mu^2}{(2-\hat\alpha^2 - \mu^2)n}+ O\left(n^{-2}\right)\right).
\end{equation}
This breakdown can be traced back to a change in the way that the temperature scales with $n$ at large $n$:
\begin{equation}
T = \frac{(2 - \hat\alpha^2 -\mu^2)n}{8 \pi} + O\left(n^{0}\right).
\end{equation}
Consequently, in order to examine the point $\hat\alpha^2+\mu^2 =2$ we must repeat our large $n$ analysis there. For $\mu=0$ and $\hat\alpha^2 =2$ the master field equation \eqref{MF eq neutral} can be solved exactly for any $n$. This generalises the analysis performed at $n=1$ in \cite{Davison:2014lua}. The additional divergence in the $\mu\neq 0$ case \eqref{eq:QNMcharged} at $2+\mu^2 = \hat\alpha^2$ coincides with the change of $n$ scaling of the mass parameter of the background solution, and occurs at a higher value of $\hat\alpha^2$ than the divergence discussed above. 

\begin{figure}[h!]
\begin{center}
\includegraphics[width=0.9\textwidth]{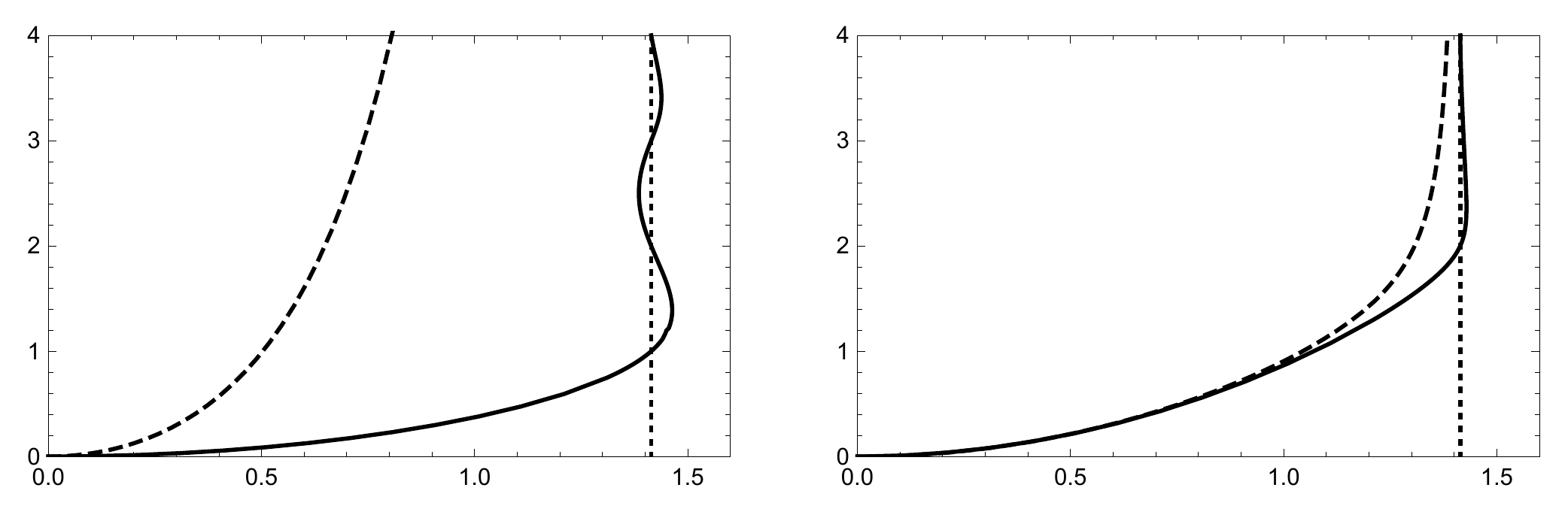}\\
\includegraphics[width=0.45\textwidth]{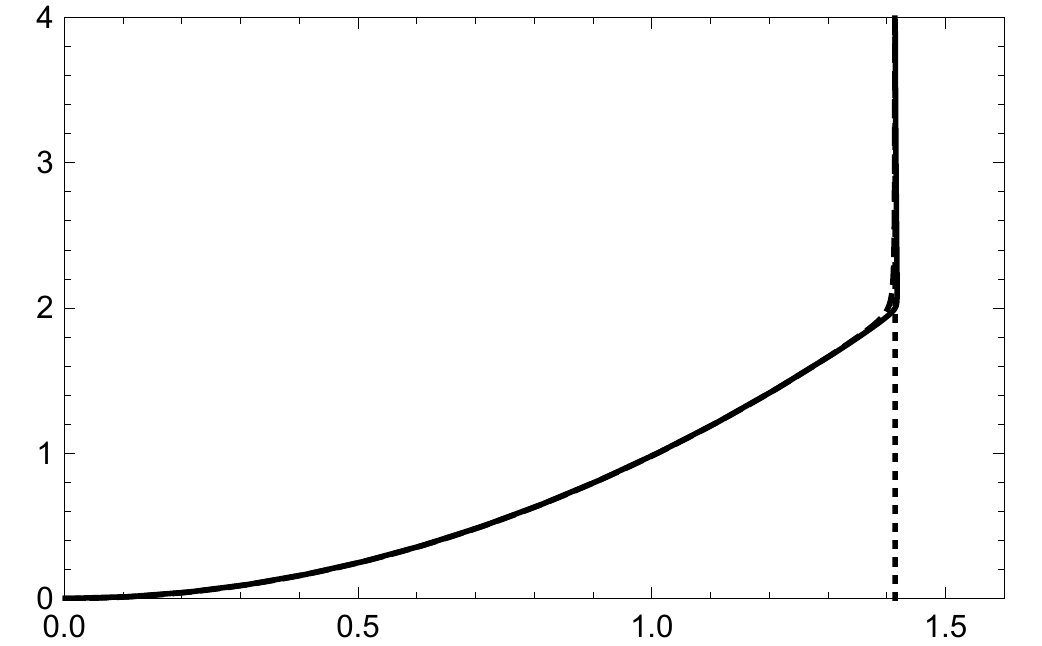}\\
\begin{picture}(0.1,0.1)(0,0)
\put(-160,240){\makebox(0,0){$n=1$}}
\put(45,240){\makebox(0,0){$n=11$}}
\put(-55,120){\makebox(0,0){$n=101$}}
\put(-215,225){\makebox(0,0){$-\Im(\omega)$}}
\put(-115,100){\makebox(0,0){$-\Im(\omega)$}}
\put(0,10){\makebox(0,0){$\hat\alpha$}}
\put(-105,135){\makebox(0,0){$\hat\alpha$}}
\put(105,135){\makebox(0,0){$\hat\alpha$}}
\end{picture}
\caption{Purely imaginary quasi-normal mode frequencies of the linear axion black brane at $\mu=0$ as a function of $\hat\alpha = \alpha/\sqrt{n}$ computed numerically for various $n$ as labelled (solid curves). The dashed lines show the large $n$ analytical counterpart \eqref{eq:QNMneutral} computed to order $n^{-2}$. The vertical dotted line is the `critical' value $\hat\alpha = \sqrt{2}$, where the master field equation can be solved analytically for any $n$, giving the integer crossing frequencies. The $n=1$ case shows several pole collisions,  indicated by each turn over of the curve, whilst for $n>1$ we see only one collision for this range. Units are given by $r_0=1$.\label{QNMplot}}
\end{center}
\end{figure}
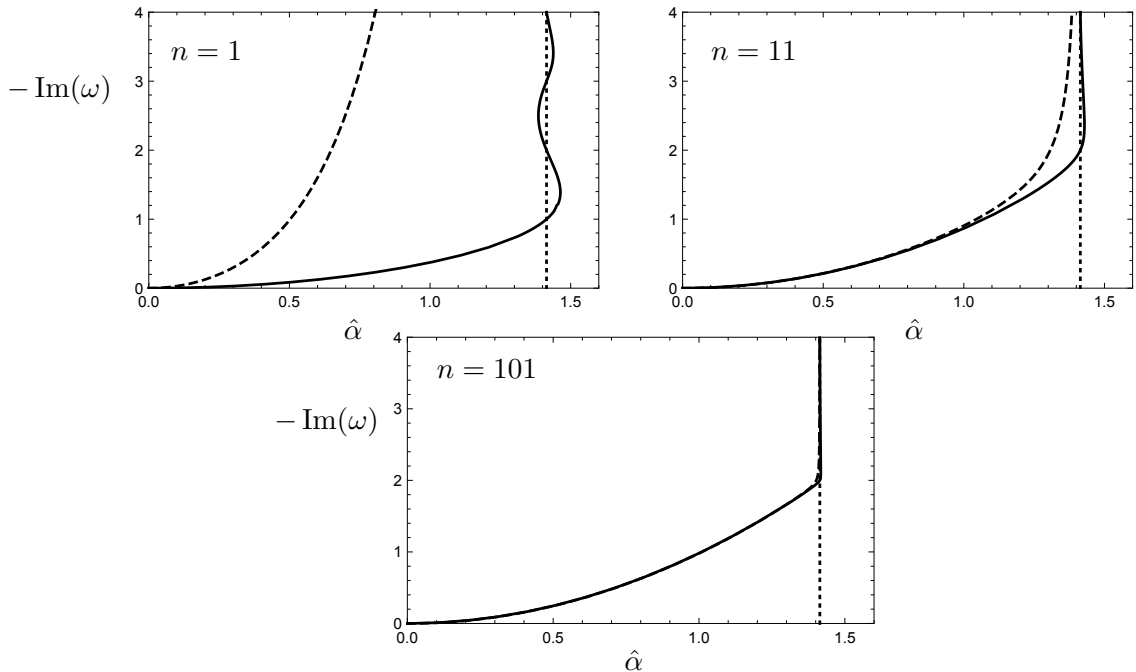

Moving on, we would like to compare these large $n$ analytical expressions \eqref{eq:QNMcharged} and \eqref{eq:QNMneutral} with finite $n$ numerics. For clarity we focus on $\mu=0$, for which the comparison is presented in figure~\ref{QNMplot} for values $n=1,11$ and $101$.  The $n=1$ case was previously analysed numerically in \cite{Davison:2014lua} wherein it was noted that a pole collision occurred as $\alpha$ was dialled. In this figure~\ref{QNMplot}, we demonstrate that there are numerous such pole collisions at $n=1$, indicated by each extremum of the curve. For $n>1$ we find only one pole collision. Interestingly, the oscillations are centred on the critical value $\hat\alpha = \sqrt{2}$ discussed above, and the locations where the crossings occur coincide with the existence of analytic regular, normalisable modes whose frequencies have integer imaginary part as discussed in appendix \ref{appendixCritical}. We discuss these collisions in the context of a transition from coherent to incoherent behaviour in the conclusions, section \ref{sec:conclusions}.

Finally, we note that there is excellent agreement at sufficiently large finite $n$ between the numerical results and the large $n$ expansion, which interestingly includes the breakdown near $\hat\alpha = \sqrt{2}$.

\section{AC conductivity}
\label{sec:conductivity}

In this section we compute the AC thermal conductivity for the neutral theory to order $n^{-2}$, and compare the resulting expressions with numerics. Specifically we look at the frequency range which captures the decoupled mode describing the essential momentum relaxation physics, i.e.\ we take $\omega = O\left(n^0\right)$.

The computation begins with the approach outlined in \cite{Emparan:2013moa}. The basic structure of the calculation is a matched asymptotic expansion made possible by the new small scale $r_0/n$, corresponding to the localisation of gradients near the horizon. This new scale allows us to separate the bulk geometry into a near and far zone, defined as follows:
\begin{eqnarray}
\text{near zone:}\quad & r-r_0\ll r_0, & \quad \log \rho \ll n\\
\text{far zone:}\quad & r-r_0\gg \frac{r_0}{n}, &\quad \log \rho \gg 1.
\end{eqnarray}
As we have previously, we shall take $r_0=1$. Note that these zones overlap: in particular, the overlap zone is described by $\log \rho \gg 1$ in the near zone, and $\log \rho \ll n$ in the far zone. Thus the calculation proceeds by solving in both zones and matching at the overlap; the near zone will allow us to imprint the ingoing horizon boundary conditions on the solution, whilst the far zone will enable us to read off the normalisable and non-normalisable data and allow the computation of the two-point function. 

\subsection{Near zone}
The near zone is reached by taking the large $n$ limit whilst working at fixed $\rho$. The calculation proceeds similarly to the QNM calculation and so we will be brief. A key difference is that we do not wish to impose normalisablity, and so the frequencies are not quantised. As before, we expand $\Phi(\rho) = \sum_{i = 0} \frac{\Phi_i(\rho)}{n^i}$, but we do not expand $\omega$. After imposing the ingoing boundary conditions we obtain, 
\begin{eqnarray}
\Phi_{0} &=& \frac{a_0}{\rho}\\
\Phi_{1} &=& \frac{2 i a_0 \omega}{(\hat\alpha^2-2)}\frac{\log{(\rho-1)}}{\rho} - \frac{2 a_0(\hat\alpha^2 - i \omega)}{\hat\alpha^2 -2}\frac{\rho-1}{\rho}
\end{eqnarray}
together with explicit expressions for $\Phi_{2}$ and $\Phi_{3}$ which we have omitted here. $a_0$ is an unconstrained  integration constant. Finally, in the overlap region, we have
\begin{eqnarray}
\Phi_{0} &=& \frac{a_0}{\rho}\\
\Phi_{1} &=& -2a_0\frac{(\hat\alpha^2-i\omega)}{\hat\alpha^2 -2} \left(1-\frac{1}{\rho}\right) + 2 a_0 \frac{i \omega}{\hat\alpha^2 -2}\frac{\log\rho}{\rho} - a_0 \frac{2 i \omega}{\hat\alpha^2-2}\frac{1}{\rho^2}.
\end{eqnarray}
where again we have evaluated the overlap expressions for $\Phi_{2}$ and $\Phi_{3}$ but we omit them here in the interest of keeping the presentation concise.

\subsection{Far zone and matching}
At leading order the far zone equations can be obtained by removing any terms which decay exponentially fast with $n$ \cite{Emparan:2013moa}. More generally, an expansion can be formed by counting powers of $r^{-n}$ in the equations of motion after inserting
\begin{equation}
\Phi = \phi + r^{-n} \psi + \ldots.
\end{equation}
Let us introduce a counting parameter $\lambda$ for this purpose, i.e. we count $\phi$ as order $\lambda^0$. In order to obtain the conductivity we need the coefficient of $r^{-n}$, and so we need to go to order $\lambda^1$.
At order $\lambda^0$ the master field equation becomes
\begin{equation}
{\cal D}\phi = 0,\qquad {\cal D} \equiv \partial_r^2 - \frac{(n-1)\hat{\alpha}^2 - 2(n+1) r^2}{2r^3 - \hat{\alpha}^2 r} \partial_r + \frac{4 \omega^2}{(\alpha^2-2r^2)^2}.  \label{lambda0}
\end{equation}
This equation can be solved explicitly in terms of Gauss hypergeometric functions,
\begin{eqnarray}
\phi &=&  \left(1-\frac{\hat{\alpha}^2}{2 r^2}\right)^{-\frac{i \omega}{\sqrt{2}\hat{\alpha}}} \Bigg(A\; _2F_1\left(-\frac{i \omega}{\sqrt{2}\hat{\alpha}}, 1- \frac{n}{2} - \frac{i \omega}{\sqrt{2}\hat{\alpha}}, 1-\frac{n}{2}, \frac{\hat{\alpha}^2}{2r^2}\right)\nonumber\\
&& + r^{-n} B\; _2F_1\left(1-\frac{i \omega}{\sqrt{2}\hat{\alpha}}, \frac{n}{2} - \frac{i \omega}{\sqrt{2}\hat{\alpha}}, 1+\frac{n}{2}, \frac{\hat{\alpha}^2}{2r^2}\right) \Bigg)\label{phifar}
\end{eqnarray}
where $A$ and $B$ are integration constants; $A$ will contribute to the non-normalisable part of $\Phi$ at infinity, $\Phi^{(0)}$, whilst $B$ will contribute to the normalisable part, $\Phi^{(n)}$. To find this in the overlap region we need to use expressions for the Gauss hypergeometric functions for large parameters. Expanding $\phi$ in powers of $n$ and similarly for the integration constants $A$ and $B$, we find, 
\begin{eqnarray}
\phi_0 &=& A_0 - \frac{2B_0}{(\hat\alpha^2 - 2)\rho}\label{far0overlap}\\
\phi_1 &=& A_1 - \frac{A_0\omega^2}{\hat\alpha^2-2} + \frac{1}{\rho} \left(-\frac{2B_1}{\hat\alpha^2-2} -2 B_0\frac{2\hat\alpha^2 + \omega^2}{(\hat\alpha^2-2)^2} -4 B_0\frac{\hat\alpha^2}{(\hat\alpha^2-2)^2} \log\rho\right)\label{far0overlapB}
\end{eqnarray}
together with similar expressions for $\phi_2$ and $\phi_3$. Subscripts denote the power of $1/n$ for which it is a coefficient.

At next order in $\lambda$, the equation for $\psi$  is sourced by $\phi$:
\begin{eqnarray}
r^n{\cal D} \left(\frac{\psi}{r^n} \right) &=& {\cal S} \label{DS}
\end{eqnarray}
where the operator ${\cal D}$ is defined in \eqref{lambda0} and where
\begin{align}
{\cal S} &= r^{-1} \frac{(2-\hat\alpha^2) \left(n \hat\alpha^2  - 2(2+n) r^2\right)}{(\hat\alpha^2-2r^2)^2} \phi' \nonumber\\
&\phantom{=\ }+ r^{-2} \frac{2-\hat\alpha^2}{2r^2-\hat\alpha^2}\left(-n(2+n) - \frac{8 \omega^2 r^2}{(2r^2-\hat\alpha^2)^2}\right)\phi.
\end{align}
Unlike for $\phi$ we have not directly integrated this equation. However, we can do so order-by-order in a large $n$ expansion at fixed $r$ provided we include the correct non-perturbative contributions. To the order of $n$ considered these turn out to be,
\begin{eqnarray}
\psi &=& \left(\psi_{B,0}(r) + \frac{\psi_{B,1}(r)}{n} + \frac{\psi_{B,2}(r)}{n^2} + \frac{\psi_{B,3}(r)}{n^3}  +O\left(n^{-4}\right) \right)\nonumber\\
&&+ r^{-n}\left(\psi_{C,0}(r) + \frac{\psi_{C,1}(r)}{n} + \frac{\psi_{C,2}(r)}{n^2} + \frac{\psi_{C,3}(r)}{n^3} + O\left(n^{-4}\right) \right). \label{nonpert}
\end{eqnarray}
We can solve for each $\psi_{B,i}$ and $\psi_{C,i}$ provided $A_0=0$, which as we shall see shortly is consistent with the required value from the matching calculation.  Each term is required for the matching calculation to work and is straightforward to obtain. This method is more efficient than solving \eqref{DS} at arbitrary $n$ and then expanding, as in \cite{Emparan:2013moa}. Applied to $\phi$ above, this method gives the same result as the expansion of \eqref{phifar}.

Expressing \eqref{nonpert} in the overlap region and combining with \eqref{far0overlap} and \eqref{far0overlapB} gives us $\Phi$ in the overlap zone, which can be matched with the expression coming from the near zone calculation. This fixes the coefficients appearing in \eqref{far0overlap},\eqref{far0overlapB} together with additional integration constants which arise in each of the $\psi_{B,i}$. For example, 
\begin{align}
A_0 &= 0,\\
A_1 &= \frac{2 a_0(\hat\alpha^2-i\omega)}{2-\hat\alpha^2}\\
A_2 &= -\frac{2 a_0(-i \omega^3 + \hat\alpha^2(4-2i\omega+\omega^2))}{(2-\hat\alpha^2)^2}\\
\nonumber
A_3 &= \frac{a_0}{3\left(\hat{\alpha }^2-2\right)^3} 
\bigg\{ \hat{\alpha }^4 \left(-6 \omega ^2-4 i \pi ^2 \omega \right) \\
   &-\hat{\alpha }^2 \left[ 3 \left(\omega ^4-6 i \omega
   ^3+8 \omega ^2+32\right)+4 \pi ^2 \omega  (\omega +2 i)\right] + i \omega ^2 \left(3 \omega ^3-4 \pi ^2 \omega +48 i\right) \bigg\}
\end{align}
The $B_i$ coefficients are given in relation to the coefficients appearing in $\psi_{B,i}$. These coefficients determine $\Phi^{(0)}$ and $\Phi^{(n)}$ to order $n^{-3}$. Note that since $\psi$ does not contribute to $\Phi^{(0)}$, and $A_0=0$, the non-normalisable data $\Phi^{(0)}$ vanishes to leading order in $n$ and so the Green's function will grow with $n$.

\subsection{Results}
Combining the asymptotic results for $\phi$ and $\psi$ discussed above brings us to the main result of this section --- the thermal conductivity \eqref{conductivityDefs} to order $n^{-2}$: 
\begin{eqnarray}
\nonumber
&& \kappa(\omega) =  2\pi \frac{2-\hat\alpha^2}{\hat\alpha^2-i \omega} \\
\nonumber
&&+\frac{4 \pi  \left(\hat\alpha^4 (2+\omega  (\omega -i))-i \hat\alpha^2 \omega ^3+2 i \omega  \left(\hat\alpha^2-i
   \omega \right)^2 \log \left(2-\hat\alpha^2\right)-2 i \omega  \log (2) \left(\hat\alpha^2-i \omega \right)^2\right)}{n
   \left(\hat\alpha^3-i \hat\alpha \omega \right)^2}\nonumber\\
\nonumber
&&-\frac{4 \pi \omega }{3 n^2 \hat{\alpha }^4 \left(\hat{\alpha }^2-2\right)^2 \left(\omega +i \hat{\alpha }^2\right)^3}\bigg\{\\
\nonumber
&& +\hat{\alpha }^6 \bigg[ \hat{\alpha }^6 (-(6 \log (2-\hat{\alpha }^2 )+\pi ^2+6-6 \log (2) )) \\
\nonumber
&& +4 \hat{\alpha }^2 (6 \log (2-\hat{\alpha }^2 )+\pi ^2+12-6 \log (2) )-48 \bigg]\\
\nonumber
&& +i \hat{\alpha }^6 \bigg[  (\hat{\alpha }^2-2 ) (3 (\hat{\alpha }^2+\pi ^2+4 ) \hat{\alpha }^2+2 (\pi ^2-6)) \\
\nonumber
&&+18 (\hat{\alpha }^4-4 ) \log (2-\hat{\alpha }^2 ) -18 (\hat{\alpha }^4-4 ) \log (2)  \bigg] \omega\\
\nonumber
&& +\hat{\alpha}^4 \bigg[ -12 \hat{\alpha }^6 (1+\log (2))+2 \hat{\alpha }^4 (\pi ^2+15 (1+\log (2))) \\
\nonumber
&& -4 \hat{\alpha }^2 (3+\pi ^2+\log (4096))
 +6 (\hat{\alpha }^2-2) (2 \hat{\alpha }^4-\hat{\alpha }^2+6) \log (2-\hat{\alpha }^2 )+72 \log (2) \bigg] \omega ^2\\
\nonumber
&& -3 i \hat{\alpha }^2 \bigg[ \hat{\alpha }^8-3 \hat{\alpha }^6 (5+\log (16))+\hat{\alpha }^4 (28+46 \log (2))-4 \hat{\alpha }^2
 (1+\log (4096)) \\
 \nonumber
&& +2 (\hat{\alpha }^2-2) (6 \hat{\alpha }^4-11 \hat{\alpha }^2+2) \log (2-\hat{\alpha }^2)+\log (256) \bigg] \omega ^3\\
\nonumber
&& +6 \hat{\alpha }^2 \left(\hat{\alpha }^2-2\right) \bigg[-\hat{\alpha }^4+6 \hat{\alpha }^2 (1+\log (2))-6 \left(\hat{\alpha }^2-2\right) \log \left(2-\hat{\alpha }^2\right)-12 \log (2)\bigg] \omega ^4\\
\nonumber
&& +3 i \bigg[ \hat{\alpha }^6-2 \hat{\alpha }^4 (3+\log (4))+8 \hat{\alpha }^2 (1+\log (4)) \\
&& +4(\hat{\alpha }^2-2)^2  \log (2-\hat{\alpha }^2 )-16 \log (2) \bigg ] \omega ^5 \bigg\} +O\left(n^{-3}\right). \label{conductivityfinal}
\end{eqnarray}
We have fixed the overall normalisation given by $\xi$ in \eqref{G2 neutral} by comparing with the DC value \eqref{kappaDC}:
\begin{equation}
\kappa(0) = 2\pi\left(\frac{2}{\hat\alpha^2}-1\right) \left(1+ \frac{4}{(2-\hat\alpha^2)n}\right). \label{kappaDCNetural} 
\end{equation}
We find  $\xi = -\hat\alpha^2+ O(n)^{-3}$ . Interestingly, at $\mu=0$, the expansion for $\kappa(0)$ truncates at order $1/n$.

\begin{figure}[h!]
\begin{center}
\includegraphics[width=0.9\textwidth]{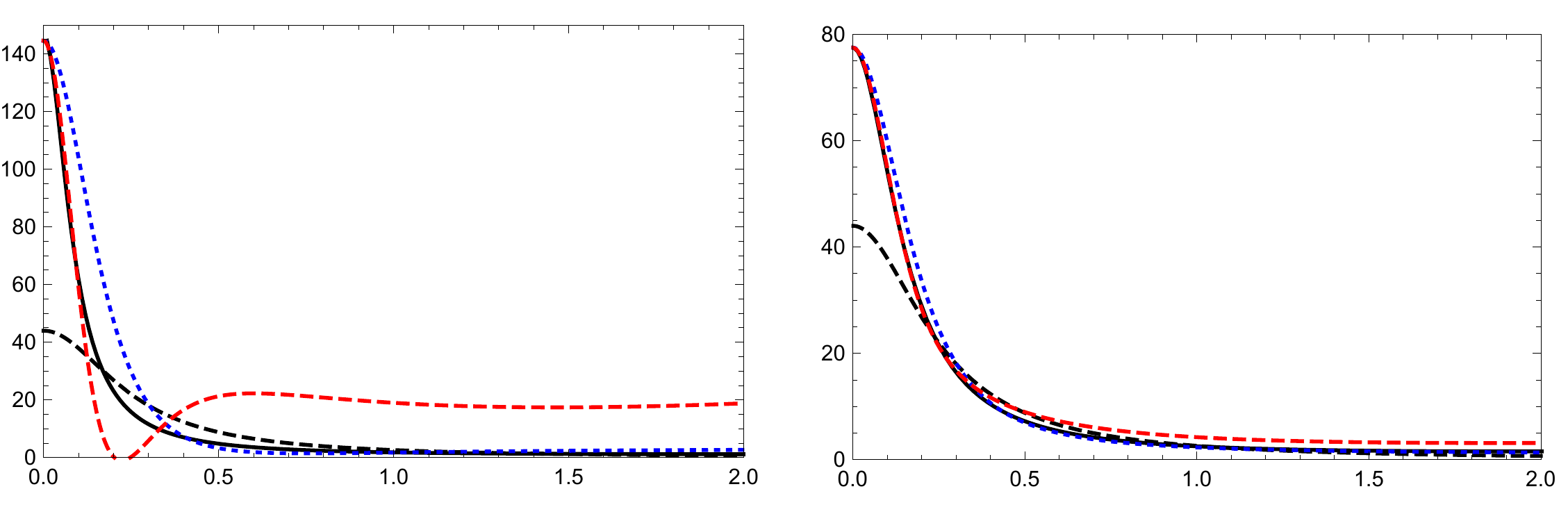}\\
\includegraphics[width=0.9\textwidth]{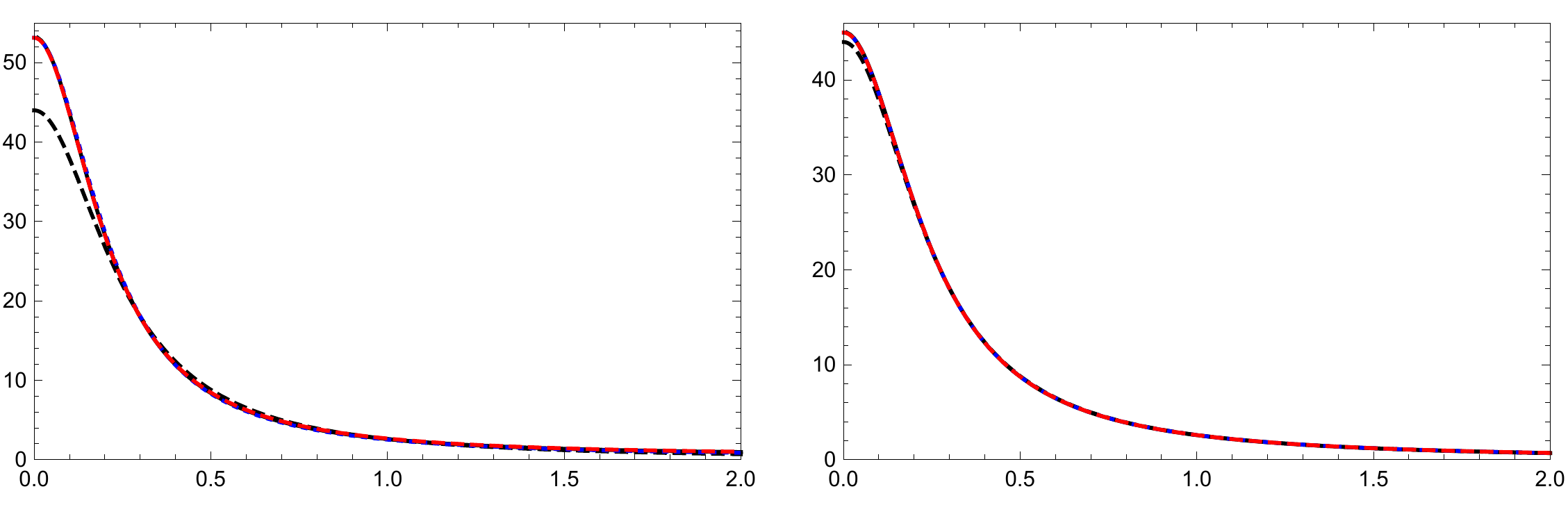}\\
\begin{picture}(0.1,0.1)(0,0)
\put(-40,250){\makebox(0,0){$n=1$}}
\put(155,250){\makebox(0,0){$n=3$}}
\put(-40,120){\makebox(0,0){$n=11$}}
\put(155,120){\makebox(0,0){$n=101$}}
\put(-210,230){\makebox(0,0){$\Re(\kappa)$}}
\put(-210,100){\makebox(0,0){$\Re(\kappa)$}}
\put(-105,10){\makebox(0,0){$\omega$}}
\put(105,10){\makebox(0,0){$\omega$}}
\end{picture}
\caption{The AC thermal conductivity at $\mu=0$ computed analytically to orders $n^0$ (black dashed), $n^{-1}$ (blue dotted) and $n^{-2}$ (red dashed) as given in \eqref{conductivityfinal}, compared to the finite $n$ numerical result (solid) for $\hat\alpha = 1/2$.  Units are given by $r_0=1$.\label{kappaplot}}
\end{center}
\end{figure}
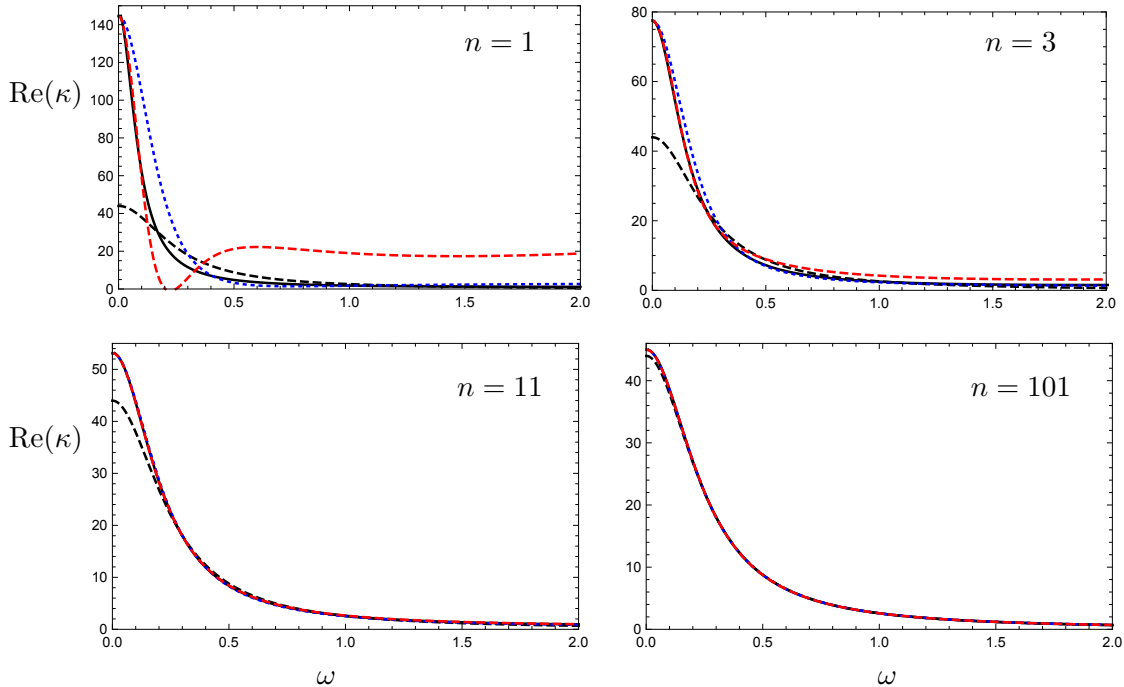

We note that at leading order the result takes Drude form.  A comparison with numerical integration for finite $n=1,3, 11,101$ is given in figure \ref{kappaplot}, truncating at orders $n^0$, $n^{-1}$ and $n^{-2}$.

Let us start with the $n^0$ approximation (black dash in figure \ref{kappaplot}).  Even at this leading order the broad features of the conductivity are well captured by the large $n$ expansion. We note that the agreement at larger frequencies is excellent. The width of the peak, which is related to the relaxation timescale, is also very good, in agreement with the approximation of the QNMs by the large $n$ expansion. The only notable discrepancy is the height of the peak. This can be easily understood: $\kappa(0)$ receives a $1/n$ correction \eqref{kappaDCNetural} and no further corrections, thus the DC limit is not expected to agree at this order. By $n=101$ the agreement is good everywhere.

At order $n^{-1}$ (blue dots in figure \ref{kappaplot}) the DC limit now agrees, as anticipated. Overall the approximation is good even at $n=1$, but now we see even for modest values of $n$ the analytical result is in excellent agreement with the numerical result, e.g.\ at $n=11$. 

At order $n^{-2}$ (red dash in figure \ref{kappaplot}) the $n=11$ and $n=101$ results are not visibly affected. However for the lower values of $n=1,3$ the agreement becomes worse than at order $n^{-1}$. This is similar to the large $D$ approximation applied to the Gregory-Laflamme instability \cite{Emparan:2015rva}, where it was argued that the series is asymptotic due to the existence of non-perturbative contributions.

\section{Conclusions}
\label{sec:conclusions}

We have studied the linear axion model defined via \eqref{S0} when the number of spacetime dimensions
is large, focusing on the quasi-normal modes governing momentum relaxation and the AC thermal conductivity. We have kept the horizon radius $r_0$ and the chemical potential $\mu$ fixed in this limit. We 
found that the influence of the momentum relaxation parameter, $\alpha$, vanished at large $n$, but could be restored by scaling $\alpha$ 
such that $\hat\alpha \equiv \alpha/ \sqrt{n}$ is held fixed. This places the physics of momentum relaxation at leading order in the large $D$ expansion.

An important technical point which simplifies our analysis is the existence of 
master field equations in arbitrary dimensions \eqref{MF eqs pm}, \eqref{MF eq neutral}. These decoupled wave equations contain 
all the gauge invariant information required to compute the two-point functions of interest, reducing our problem 
to calculations closely related to those already carried out in the context of General Relativity in large $D$. 

We obtain analytical expressions for the QNM which control the electric and heat transport as a power series in $1/D$. 
In the language of \cite{Emparan:2014aba}, these are of the decoupled kind, meaning that they are normalisable 
in the near horizon geometry. 
For $\mu \neq 0$, we have computed the QNM which controls the electric conductivity for   
up to order $n^{-1}$, while for $\mu = 0$ we obtain the thermal QNM to order $n^{-3}$. Furthermore, 
we calculate the AC thermal conductivity for $\mu = 0$ to order $n^{-2}$. At leading order it takes Drude form, 
and at order $n^{-1}$ it provides a good approximation even for small values of $n$, illustrating the practicality of this technique for such systems.

Interestingly, our perturbative series for the QNM breaks down due to the growth of the 
coefficients as $\hat\alpha \to  \hat \alpha_{c} \equiv  \sqrt{2 r_0^2-\mu^2}$. 
Numerically, we observe that the structure of lowest lying QNM 
changes significantly as we approach $\hat \alpha_{c}$. For very small values of $\hat \alpha$ there exists an isolated, 
purely dissipative excitation which governs transport, i.e.\ the system is in a coherent regime. Increasing $\alpha$ towards 
$\hat \alpha_{c}$, the characteristic time scale of this mode decreases and it mixes with the rest of the QNM in the spectrum, so 
that we enter  an incoherent phase.\footnote{In this regime there can be numerous pole collisions at finite $n$, including a collision between the Drude mode and a higher lying excitation.}
We thus interpret the breakdown of the perturbative expansion as a large $D$ signature 
of the coherent/incoherent transition. It would be interesting to revisit our analysis with transverse wavevector $k \neq 0$ and investigate the interplay of these QNMs with diffusion.

More generally, it is interesting to observe that in the case where we do not scale the sources of the axions with $D$, momentum conservation is restored at infinite $D$. This suggests that the large $D$ expansion may be used to improve analytical control over more generic setups incorporating inhomogeneity. We leave this possibility for future work.

\acknowledgments

We are pleased to thank Marco Caldarelli, Richard Davison, Roberto Emparan, Blaise Gout\'eraux and Kostas Skenderis for valuable comments. 
T.A.\ is supported by the European Research Council under the European Union's Seventh Framework Programme
(ERC Grant agreement 307955). He also thanks the Institute of Physics at University of Amsterdam for their hospitality 
during the completion of this work. 
S.A.G.\ is supported by National Science Foundation grant PHY-13-13986.
B.W.\ is supported by European Research Council grant ERC-2014-StG639022-NewNGR.  He also thanks the Department of Physics at the University of Oxford for their hospitality during the completion of this work.
We are grateful to Centro de Ciencias Pedro Pascual, Benasque where this work was initiated.

\appendix

\section{Critical modes}\label{appendixCritical}
In this appendix we focus on the critical value of  $\hat\alpha_c = \sqrt{2}$ when  $\mu=0$.  In this case the equation for the neutral master field~\eqref{MF eq neutral} takes the following simple form:
\begin{equation}\label{MF eq neutral critical}
\Phi''+\left(\frac{2r}{r^2-1}+\frac{n-1}{r}\right) \Phi' + \frac{\omega^2}{\left(r^2-1\right)^2}\, \Phi = 0.
\end{equation}

It is immediately clear that $n=1$ is a special case.  For generic $\omega$ we cannot simultaneously impose  normalisability at infinity as well as kill the outgoing mode at the horizon.  However, when $ \omega = -i m$ we find the following special solution:
\begin{equation}
\Phi_c = A \left(r^2-1\right)^{-m/2} \left[\left(r+1\right)^{m}-\left(r-1\right)^{m}\right], \quad \omega = - mi, \quad m= 1,2,\ldots,\quad n=1
\end{equation}
This is normalisable at infinity and also, upon switching to  ingoing Eddington-Finkelstein coordinates,   regular at $r=1$.  When $n>1$ the following  solution satisfies these boundary conditions:
\begin{equation}
\Phi_c = A \, \frac{r^{2 - n}}{ r^2-1}, \quad \omega = - 2i, \quad n\geq 1
\end{equation}
These special frequencies appear in figure~\ref{QNMplot} as the points at which the purely imaginary QNM crosses the  $\hat\alpha =\hat\alpha_c$ line.

The following formula yields the AC thermal conductivity for the critical value of $\hat\alpha_c = \sqrt{2}$ at arbitrary odd $n$:
\begin{equation}
\kappa_c(\omega)=\frac{2 \pi  \cosh \left(\frac{\pi  \omega }{2}\right) \Gamma \left(\frac{1}{2} (n-i \omega )\right) \Gamma \left(\frac{1}{2} (n+i \omega
   )\right)}{\Gamma \left(\frac{n}{2}+1\right) \Gamma \left(\frac{n}{2}\right)}, \quad n\ \textrm{odd}
\end{equation}
extracted using  \eqref{conductivityDefs}, \eqref{G2 neutral}.  It reduces to a constant for $n=1$, but otherwise is frequency-dependent. In particular, this expression has no poles as a function of  $\omega$.    This generalises the analysis of  \cite{Davison:2014lua}.  

\bibliographystyle{JHEP-2}
\bibliography{drude}

\end{document}